\documentclass[reprint,twocolumn,superscriptaddress,secnumarabic,amssymb,nobibnotes,aps,prx]{revtex4-2}
\usepackage{graphicx}
\usepackage{dcolumn}
\usepackage{bm}
\usepackage{xcolor}
\usepackage[export]{adjustbox}
\usepackage{float}
\usepackage{tabularx}
\usepackage{multirow}
\usepackage{amsmath}
\usepackage{braket}
\usepackage{array}
\usepackage{ulem}
\usepackage{comment}

\newcolumntype{C}[1]{>{\centering\let\newline\\\arraybackslash\hspace{0pt}}m{#1}}

\begin{document}
	
	
	\title{UCd$_{11}$: a strongly localized 5$f^3$ material}

	\author{Martin Sundermann}
	\altaffiliation{These authors contributed equally to this work.}
	\affiliation{Max Planck Institute for Chemical Physics of Solids, N{\"o}thnitzer Stra{\ss}e 40, 01187 Dresden, Germany}
	\affiliation{PETRA III, Deutsches Elektronen-Synchrotron DESY, Notkestra{\ss}e 85, 22607 Hamburg, Germany}
	
	\author{Naoki Ito}
	\altaffiliation{These authors contributed equally to this work.}
	\affiliation{Department of Physics and Electronics, Osaka Metropolitan University 1-1 Gakuen-cho, Nakaku, Sakai, Osaka 599-8531, Japan} 
	
	\author{Daisuke~Takegami}
	\altaffiliation{Present address: Department of Physics, Tokyo Metropolitan University, Hachioji, 192-0397, Japan.}
	\affiliation{Max Planck Institute for Chemical Physics of Solids, N{\"o}thnitzer Stra{\ss}e 40, 01187 Dresden, Germany}
	
	\author{Chun-Fu~Chang}
	\affiliation{Max Planck Institute for Chemical Physics of Solids, N{\"o}thnitzer Stra{\ss}e 40, 01187 Dresden, Germany}
	
	\author{Sheng-Huai~Chen}
	\affiliation{Max Planck Institute for Chemical Physics of Solids, N{\"o}thnitzer Stra{\ss}e 40, 01187 Dresden, Germany}
	
	\author{Chang-Yang~Kuo}
	\affiliation{Department of Electrophysics, National Yang Ming Chiao Tung University, Hsinchu, 30010, Taiwan  }
	\affiliation{National Synchrotron Radiation Research Center, 101 Hsin-Ann Road, Hsinchu 300092, Taiwan  }
	
	\author{Simone~G.~Altendorf}
	\affiliation{Max Planck Institute for Chemical Physics of Solids, N{\"o}thnitzer Stra{\ss}e 40, 01187 Dresden, Germany}
	
	\author{Andrei~Gloskovskii}
	\affiliation{PETRA III, Deutsches Elektronen-Synchrotron DESY, Notkestra{\ss}e 85, 22607 Hamburg, Germany}
	
	\author{Hlynur~Gretarsson}
	\affiliation{PETRA III, Deutsches Elektronen-Synchrotron DESY, Notkestra{\ss}e 85, 22607 Hamburg, Germany}
	\affiliation{Max Planck Institute for Solid State Research, Heisenbergstra{\ss}e 1, 70569 Stuttgart, Germany}
	
	\author{Eric~D.~Bauer}
	\affiliation{Los Alamos National Laboratory, Los Alamos, New Mexico 87545, USA}
	
	\author{Shin-ichi~Fujimori}
	\affiliation{Materials Sciences Research Center, Japan Atomic Energy Agency, Sayo, Hyogo 679-5148, Japan}
	
	\author{Jan~Kune{\v s}}
	\affiliation{Department of Condensed Matter Physics, Faculty of Science, Masaryk University, Kotl\'a\v{r}sk\'a 2, 611 37 Brno, Czechia}
	
	\author{Liu~Hao~Tjeng}
	\affiliation{Max Planck Institute for Chemical Physics of Solids, N{\"o}thnitzer Stra{\ss}e 40, 01187 Dresden, Germany}
	
	\author{Andrea~Severing}
	\altaffiliation{Contact author: andrea.severing@cpfs.mpg.de}
	\affiliation{Max Planck Institute for Chemical Physics of Solids, N{\"o}thnitzer Stra{\ss}e 40, 01187 Dresden, Germany}
	\affiliation{Institute of Physics II, University of Cologne, Z\"{u}lpicher Stra{\ss}e 77, 50937 Cologne, Germany}
	
	\author{Atsushi~Hariki}
	\altaffiliation{Contact author: hariki@omu.ac.jp}
	\affiliation{Department of Physics and Electronics, Osaka Metropolitan University 1-1 Gakuen-cho, Nakaku, Sakai, Osaka 599-8531, Japan}
	\date{\today}

	\begin{abstract}
			UCd$_{11}$ is an antiferromagnetic uranium intermetallic compound ($T_{\rm N}$\,=\,5.3\,K) with enhanced electron mass and uranium--uranium spacings nearly twice the Hill limit, suggesting a weakly hybridized 5$f$ electronic character. Various x-ray spectroscopy techniques indicate that uranium in UCd$_{11}$ adopts the formal U$^{3+}$\,5$f^3$ configuration, while core-level photoemission spectroscopy (PES) data of UCd$_{11}$ reveal only a weak satellite feature, typically interpreted as a signature of itinerancy. In this work, we present density functional theory (DFT) combined with dynamical mean-field theory (DMFT) calculations of UCd$_{11}$, using material-specific parameters tuned to reproduce valence-band PES spectra at different photon energies, thereby exploiting the energy dependence of photoionization cross sections. Our results demonstrate that UCd$_{11}$ is a highly localized uranium 5$f^3$ system. Furthermore, core-level spectra obtained from a DFT\,+\,DMFT Anderson impurity model reveal that, contrary to common assumptions, the presence or absence of satellite structures is not a reliable indicator of strong correlations or itinerant 5$f$ behavior.
	\end{abstract} 
	
	\maketitle

	\section{Introduction}
    Establishing a reliable electronic model is an essential step toward understanding the driving mechanisms behind unusual physical phenomena, such as unconventional superconductivity and magnetism, in strongly correlated materials. In such systems, the formal valency, the degree of covalency of the constituent elements and their orbitals, and the strength of electronic correlations are key parameters. However, for uranium intermetallic compounds, this information is particularly difficult to obtain experimentally because the local U\,5$f$ electronic correlations are of the same order of magnitude as the hybridization of the U\,5$f$ electrons with the surrounding electron bath.  This complicates the interpretation even of relatively simple spectroscopic techniques such as photoemission and x-ray absorption.
    Standard density functional theory (DFT) is usually insufficient to describe their correlated ground state.

    DFT combined with dynamical mean-field theory (DFT+DMFT) was introduced in the late 1990s~\cite{Georges1996,kotliar06} and has been applied to  actinide compounds, including light actinide oxides~\cite{Koloren2015} as well as intermetallic systems such as URu$_2$Si$_2$~\cite{Haule2009}, UTe$_2$~\cite{Miao2020,Xu2019,Choi2024,Kang2024}, and UGa$_2$~\cite{Chatterjee2021}.
    However, ambiguities may arise from the choice of model parameters, in particular the so-called double-counting correction $\mu_{\rm dc}$~\cite{kotliar06,Karolak10}, which can yield different U\,5$f$ occupancies depending on its value. Recently, an approach has been proposed to largely reduce this ambiguity in DFT+DMFT simulations of uranium intermetallic compounds by determining material-specific parameters that reproduce experimental photon-energy-dependent photoemission spectra.
    At first, the two model compounds, ferromagnetic UGa$_2$ ($T_C$\,=\,124\,K) and paramagnetic UB$_2$, representing opposite extremes of localization and delocalization\,\cite{Marino2024}, and later the spin triplet superconductor UTe$_2$\,\cite{SundermannUTe2} were treated within this framework. In all three compounds, the U\,5$f^2$ configuration was found to be the most prevalent, with UGa$_2$ and UTe$_2$ being strongly correlated and UB$_2$ strongly itinerant, close to the bandlike limit.

	
	\begin{figure*}[t]
		\begin{center}
			\includegraphics[width=1.99\columnwidth]{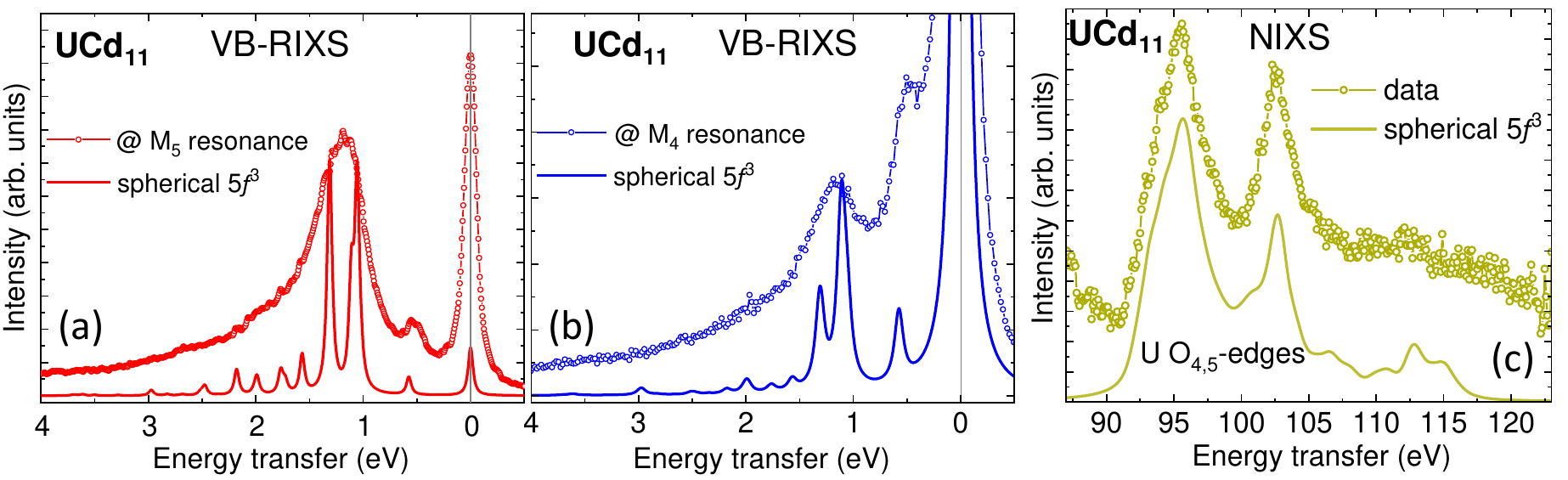}
		\end{center}
		\caption{Reported x-ray scattering spectra of UCd$_{11}$ (open circles) and corresponding spherical full-multiplet calculations based on the U\,5$f^3$ configuration (lines): (a), (b) VB-RIXS at the U\,\textit{M}$_5$ (red) and U\,\textit{M}$_4$ edges (blue), respectively; (c) NIXS at the U\,\textit{O}$_{4,5}$ edges (dark yellow). Data are adapted from Refs.~\cite{Amorese2020,Christovam2024} and are provided for reader's convenience.}
		\label{oldDATA}
	\end{figure*}
	
	Here we turn to UCd$_{11}$. This material orders antiferromagnetically at $T_N$\,=\,5.3\,K, and its magnetic susceptibility follows a Curie law with an effective moment of 3.38\,$\mu_B$~\cite{Yamamoto2012}. The compound crystallizes in the cubic cagelike BaHg$_{11}$ structure (\textit{Pm$\bar{3}$m}), with the U ions occupying the 3$c$ sites. These sites possess tetragonal symmetry, resulting in three equivalent U positions per unit cell, each rotated by 90$^{\circ}$ relative to the others. Each U atom is surrounded by 12 nearest- and 8 next-nearest-neighbor Cd ions, making the unit cell large. The U–U distance is very large, $d_{UU}=6.56$\,\AA~\cite{Cafasso1963,Fisk1984,Zaremba2022}, well above the Hill limit of $\approx 3.5$\,\AA, making direct overlap of the 5$f$ orbitals negligible. Among U intermetallics, such large U–U separations are otherwise observed only in the face-centered cubic, cagelike structure of the superconductor UBe$_{13}$ ($Fm\bar{3}c$, $d_{UU}=5.13$\,\AA)~\cite{McElfresh1990}.
	UCd$_{11}$ is further characterized by a large electronic specific heat in the paramagnetic state, initially reported to be as high as $\gamma\approx800$\,mJ/mol\,K$^2$, which decreases to about 250\,mJ/mol\,K$^2$ below the magnetic transition~\cite{Fisk1984}. Later de Haas-van Alphen (dHvA) work revealed similarly small Fermi surfaces in both UCd$_{11}$ and ThCd$_{11}$, suggesting that the enhanced $\gamma$ value originates from magnetic contributions to the specific heat rather than from heavy quasiparticles. These findings further reduced the estimated $\gamma$ to about 100\,mJ/mol\,K$^2$~\cite{Hirose2013}.

	Thus, if the Doniach phase diagram\,\cite{Doniach1977} is to be used, UCd$_{11}$ may best be described as being in the \textit{local-moment} regime of the Doniach diagram, with relatively weak $f$-electron/conduction electron hybridization.  This scenario is consistent with low-temperature neutron diffraction results, which place an upper bound on the ordered moment of 1.5\,$\mu_B$/U\cite{Thompson1988}.	
	
	UCd$_{11}$ appears to be one of the few intermetallic uranium compounds in which the uranium ions adopt the dominant 5$f^3$ configuration. However, previous studies have shown that experimentally confirming the 5$f^3$ character, as opposed to a 5$f^2$ configuration, is not straightforward. The effective magnetic moments obtained from the slope of the temperature dependence of the inverse static susceptibility are very similar (3.58 and 3.62~$\mu_B$, respectively). Since the ground-state multiplets possess similar multiplicities, the information that can be derived from specific-heat data is also limited. Nevertheless, an entropy of $R\ln 2$ observed around the ordering temperature in the specific heat~\cite{Yamamoto2012}, together with dHvA results~\cite{Hirose2013}, is consistent with a doublet crystal-electric field 5$f$ ground state, suggestive of a Kramers doublet ground state, which in turn implies a 5$f^3$ configuration.
	
	The interpretation of x-ray spectroscopy data for U intermetallics is often problematic, leading to discrepancies in determining the dominant $5f$ configuration, as shown by conflicting results for compounds such as UTe$_2$ and URu$_2$Si$_2$ (see Ref.~\cite{Christovam2024} and references therein). These discrepancies arise because the interplay of correlations, covalency, and core-hole effects manifests differently in each spectroscopic process~\cite{Marino2024}. For UCd$_{11}$, several x-ray spectroscopies indicate a dominant 5$f^3$ character: the 5$f^n$ spectral-weight analysis in $L$-edge resonant x-ray emission spectroscopy (RXES)~\cite{Booth2012,Soderlind2016}, the white-line position in partial fluorescence yield (PFY) $L_3$ absorption~\cite{Booth2016}, the branching ratio in U \textit{M}-edge high-energy-resolution fluorescence detection (HERFD)~\cite{Tobin2019}, and the line shape in non-resonant inelastic x-ray scattering (NIXS)~\cite{Amorese2020} as well as in valence-band resonant inelastic x-ray scattering (VB-RIXS)~\cite{Christovam2024}. Figures~\ref{oldDATA}(a)–~\ref{oldDATA}(c) show the VB-RIXS and NIXS spectra of UCd$_{11}$ (open circles) together with spherical full-multiplet calculations (solid lines) based on a 5$f^3$ ionic configuration, adapted from Refs.~\cite{Christovam2024,Amorese2020}.
	
\begin{figure}[t]
		\begin{center}
			\includegraphics[width=0.95\columnwidth]{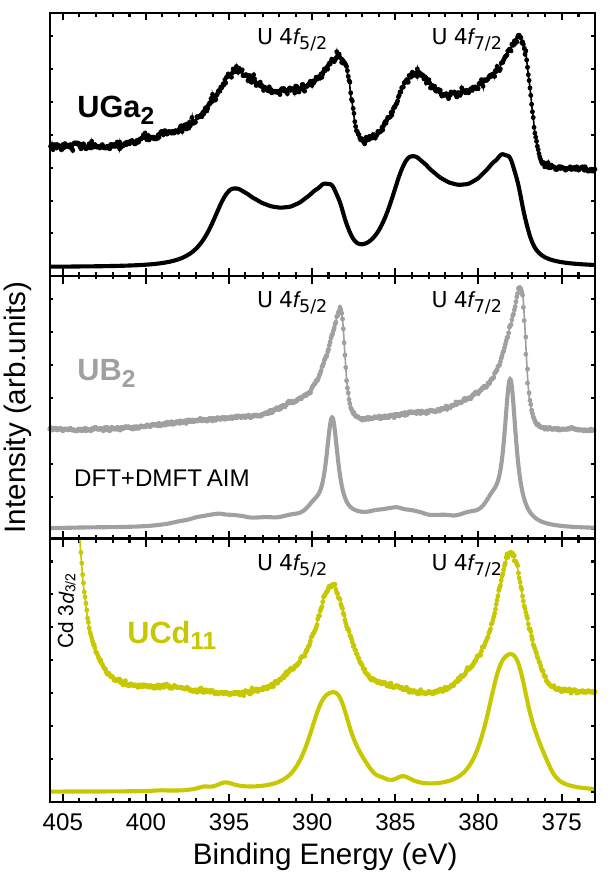}
		\end{center}
		\caption{Integral-type ("Shirley") background corrected U\,4$f$ core-level spectra of UGa$_{2}$ (black), UB$_2$ (grey), and UCd$_{11}$ (dark yellow) compared to DFT\,+\,DMFT AIM calculations. The results for UGa$_2$ and UB$_2$ are taken from previous work~\cite{Marino2024}. The small feature in the UCd$_{11}$ spectrum at 398eV is due to a plasmon excitation (see Appendix).}
		\label{core_level}
	\end{figure}

	However, questions arise when comparing the U\,4$f$ core-level photoemission spectroscopy (PES) data of UCd$_{11}$ with those of itinerant UB$_2$, as summarized in Fig.~\ref{core_level}, with data taken from Refs.~\cite{Amorese2020,Marino2024}. Besides a shift of about 0.5\,eV to higher binding energy, the main emission line in UCd$_{11}$ is broader than that of UB$_2$. Most attention, however, has been paid to the presence or absence of satellites: strong satellites, as observed in UGa$_2$, are usually interpreted as evidence of strong correlations, whereas weak satellites are taken as indicative of itinerant U 5$f$ electrons (see, e.g., Ref.~\cite{Fujimori2015}). Both UCd$_{11}$ and UB$_2$ exhibit only weak satellites; by this criterion, the 5$f$ electrons in UCd$_{11}$ would appear itinerant. 
	This view, however, stands in stark contrast to the aforementioned spectroscopic studies, in particular the multiplet excitations observed in VB-RIXS and their successful description by 5$f^3$ ionic full-multiplet calculations [Figs.~\ref{oldDATA}(a) and ~\ref{oldDATA}(b)].

    This inconsistency highlights a more general issue. Core-level PES, a standard technique for investigation of correlated materials that has been routinely applied to rare-earth and transition-metal compounds, proves far more troublesome in uranium intermetallics, where the interpretation of the spectra remains ambiguous. Here, we address this issue by means of DFT\,+\,DMFT approach.    We constrain the electronic model of UCd$_{11}$ within DFT\,+\,DMFT by VB-PES data measured with both soft and hard x-rays. Following Refs.~\cite{Christovam2024,Marino2024,SundermannUTe2}, we exploit the energy dependence of the photoionization cross-sections to identify the orbital contributions in the VB spectra near the Fermi energy $E_F$, which allows us to eliminate the model parameters in the DFT\,+\,DMFT approach. With this model, we show that UCd$_{11}$ is a strongly correlated, strongly localized U\,5$f^3$ compound. We demonstrate that the absence of satellites in the U\,4$f$ core-level spectra is not inconsistent with the localized character of the 5$f$ electrons. Finally, a simple model analysis is provided to clarify schematically the material dependence of the satellite intensity.
	
	\section{Experiment}
	UCd$_{11}$ single crystals were grown as described in Ref.~\cite{Amorese2020}. The soft x-ray VB-PES data with 600\,eV incident energy, 60\,meV resolution and 10\,meV step width were collected at the NSRRC-MPI TPS 45A Submicron Soft X-ray Spectroscopy beamline\,\cite{huangming2019} at the Taiwan Photon Source with the analyzer in the horizontal plane at 60$^{\circ}$ and sample emission along the surface normal. The hard x-rays VB-PES data with 6000\,eV incident energy and 230\,meV resolution were measured at P22 beamline\,\cite{schlueter2019} at PETRA III (DESY) in Hamburg, Germany with the analyzer in the horizontal plane at 90$^{\circ}$ from the incoming beam and with the sample surface 45$^{\circ}$ from the incoming beam. The soft and hard VB-PES data were taken at 20\,K and 40\,K, respectively. Clean sample surfaces were achieved by cleaving \textit{in situ} in ultra-high-vacuum. The pressures of the main chambers were in the low 10$^{-10}$\,mbar range. Valence band spectra of a gold or silver foil were used for calibrating the Fermi level and for determining the overall resolutions.
	
	

	\begin{figure*}[t]
		\begin{center}
			\includegraphics[width=1.99\columnwidth]{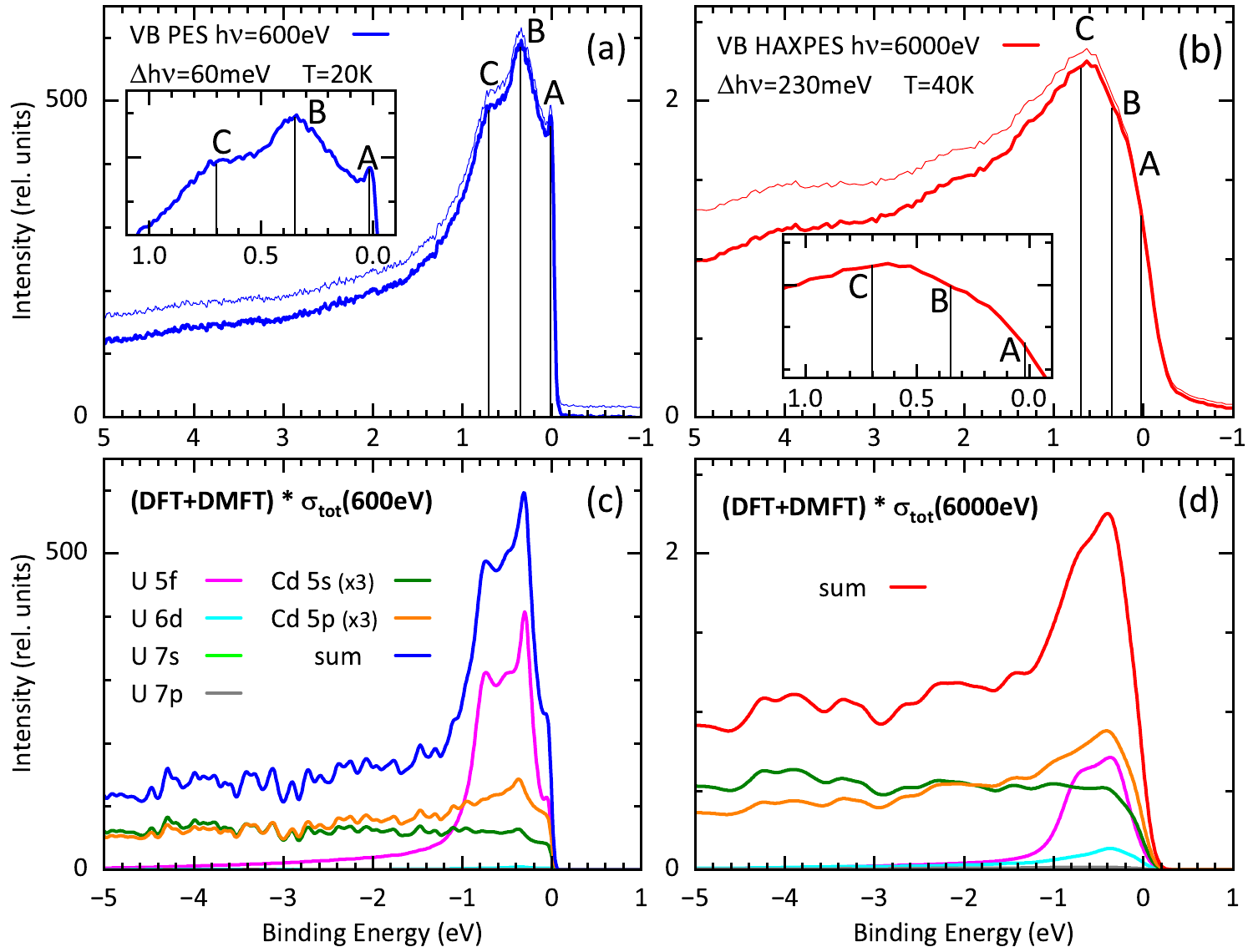}
		\end{center}
		\caption{VB-PES spectra of UCd$_{11}$ measured with soft [panel (a)] and hard [panel (b)] x-rays. The thin lines show data as measured and the thick lines after the subtraction of an integral-type ("Shirley") background (see Appendix). Panels (c) and (d) show the corresponding DFT\,+\,DMFT calculations of the total and subshell contributions, obtained with $U_{\rm ff} = 3$\,eV, $J = 0.59$\,eV, and the optimized $\mu_{\rm dc}$\,=\,5.45\,eV. The calculated spectra were broadened, multiplied by the Fermi function, and weighted using the photoionization cross sections from Table \,I.}
		\label{data}
	\end{figure*}
	
	\section{Theory} 
	The DFT+DMFT calculation for UCd$_{11}$ was performed using the same implementation as recently applied to several uranium compounds in Refs.~\cite{Marino2024,SundermannUTe2}. A standard DFT calculation for the experimental structure of UCd$_{11}$ was performed within the local density approximation for the exchange-correlation functional, using the \textsc{Wien2k} package~\cite{wien2k}. A tight-binding lattice model was then derived from the DFT bands, explicitly including U\,5$f$, 7$s$, 6$d$, 7$p$, and Cd\,5$s$ and 5$p$ states~\cite{wien2wannier,wannier90}. The model was supplemented with local Coulomb interactions among the U\,5$f$ electrons and solved within the DMFT framework~\cite{Georges1996,kotliar06}. We employed the same interaction parameters, namely Hubbard $U_{\rm ff}$\,=\,3.0\,eV and Hund's $J$\,=\,0.59\,eV, as in previous studies of UGa$_2$, UB$_2$\,\cite{Marino2024}, and UTe$_2$\,\cite{SundermannUTe2}. The U\,5$f$-electron self-energy was obtained from an Anderson impurity model (AIM) solved using the continuous-time quantum Monte Carlo method~\cite{werner06,gull11,boehnke11,hafermann12}, retaining only the density-density terms of the Coulomb vertex for computational efficiency. 
    We adopt a one-electron basis that diagonalizes the local one-particle Hamiltonian, including both spin--orbit coupling and crystal-field contributions. The one-particle Hamiltonian at the U site derived from the DFT bands is provided in Eq.~\eqref{eq:localH} of Appendix. In the core-level PES simulations described below, all Coulomb vertices for the $f$-$f$ interaction are included explicitly.
    Upon convergence of the DMFT self-consistency loop, the valence-band spectrum was calculated via analytic continuation of the self-energy using the maximum entropy method~\cite{jarrell96}. In this work, the DFT+DMFT simulations are performed for the paramagnetic phase, and the simulation temperature is set to 300~K.
	
	The core-level PES spectra were computed from the AIM using the converged DFT\,+\,DMFT hybridization densities. 
    In this step, the AIM is extended to include the core orbitals and their interaction with the valence electrons.
    In this work, the orbital degrees of freedom of the core are not treated explicitly for computational efficiency.
    The 4$f$ spectrum was generated by incorporating the degeneracies of the 4$f_{7/2}$ and 4$f_{5/2}$ states, with their energy splitting due to spin-orbit coupling taken into account. This approach has been shown to be reasonable for uranium compounds~\cite{Marino2024,SundermannUTe2}. Within this approximation, the core--valence interaction includes only the monopole term, parameterized by the core-hole potential $U_{\rm fc}$, and multiplet effects are neglected. This parameter is fixed to $U_{\rm fc}=5.0$\,eV, consistent with previous studies of uranium compounds, as its material dependence is expected to be weak. 
    When evaluating the spectral intensities, a configuration-interaction solver is adopted, in which the DFT\,+\,DMFT hybridization densities are represented by 24 discrete bath levels per spin and orbital.
    The computational procedures used for these spectra are the same as those described in Refs.~\cite{Marino2024,Hariki17}. 


	In the DFT+DMFT modeling of U compounds, determining the double-counting correction $\mu_{\rm dc}$ is a crucial step~\cite{Marino2024,SundermannUTe2}. It is introduced to compensate for the $f$–$f$ interaction effects that are already included at the DFT level~\cite{kotliar06,Karolak10,Haule15}. In practice, the on-site energies of the U 5$f$ orbitals in the Hamiltonian solved by DMFT are obtained by shifting the corresponding DFT values by $\mu_{\rm dc}$, and thus the choice of $\mu_{\rm dc}$ directly affects the uranium valence. Although no universally accepted formula for $\mu_{\rm dc}$ exists, it controls the relative energy positions between the correlated U 5$f$ states and the uncorrelated bands. A physically reasonable value for $\mu_{\rm dc}$ can be obtained by tuning it so that the DFT\,+\,DMFT calculation reproduces the experimental valence-band PES spectra~\cite{Marino2024,SundermannUTe2}, measured at different photon energies that probe the correlated 5$f$ and uncorrelated non-5$f$ states with varying sensitivity. In addition, the calculated core-level PES spectra are compared with experiment, providing complementary validation of the chosen $\mu_{\rm dc}$.

	\section{Results and Discussion}
	
	\subsection{Valence band and DFT+DMFT modeling}
	\begin{figure}[t]
		\begin{center}
			\includegraphics[width=0.99\columnwidth]{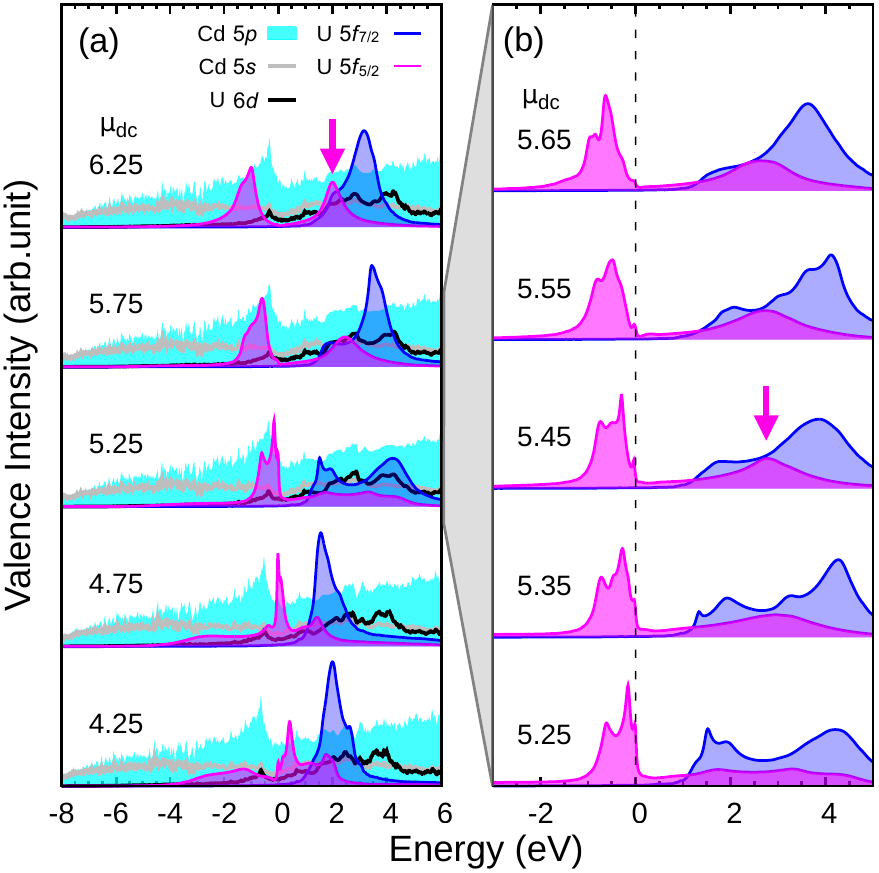}
		\end{center}
		\caption{Orbital-resolved DFT+DMFT spectral densities computed with $U_{\rm ff}=3$ eV and $J=0.59$ eV. Panel (a) shows results for $\mu_{\rm dc}$ values ranging from 6.25 to 4.25 eV in steps of 0.5 eV, while panel (b) presents a finer mesh from 5.25 to 5.65\,eV in steps of 0.1\,eV. In panel (b), only the U\,5$f$ contributions are displayed. The upper Hubbard feature is marked by a pink arrow. The dashed line indicates the Fermi level $E_F$.}
		\label{vb_mu_dep}
	\end{figure}
	
	
	\begin{figure*}[]
		\begin{center}
			\includegraphics[width=2.0\columnwidth]{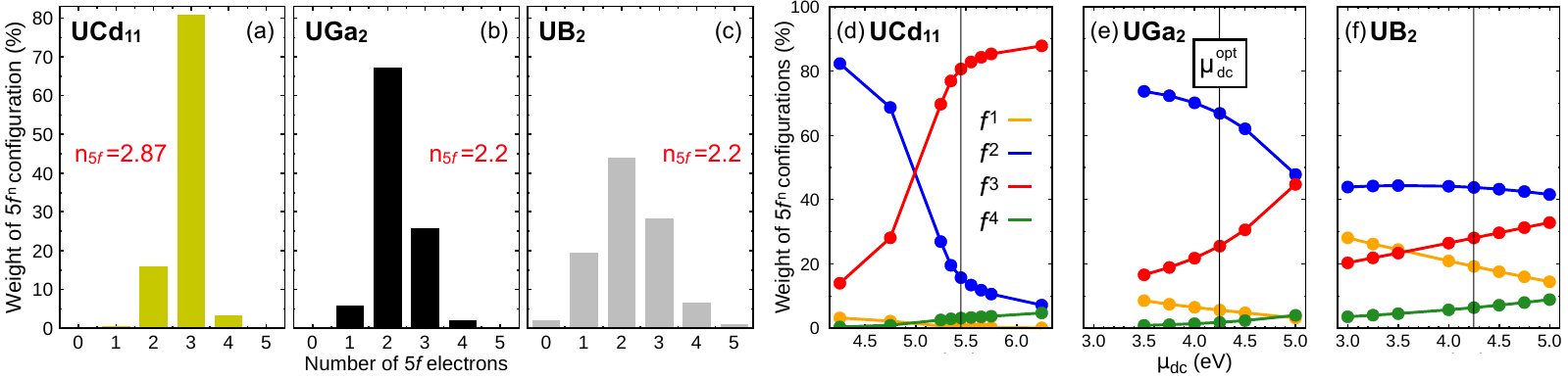}
		\end{center}
		\caption{(a)–(c) Weights of the 5$f^n$ configurations in the ground state of UCd$_{11}$, UGa$_2$ and UB$_2$, obtained from DFT+DMFT calculations with optimized parameters. 	The data for UGa$_2$ and UB$_2$ are adapted from Ref.~\cite{Marino2024}. (d)-(f) Variation of the U 5$f^n$ contributions in UCd$_{11}$, UGa$_2$ and UB$_2$ as a function of $\mu_{\rm dc}$. The respective optimal $\mu_{\rm dc}$ values are indicated by vertical lines.}
		\label{hist} 
	\end{figure*}

    Figure\,\ref{data}(a) and \ref{data}(b) show the soft x-ray ($h\nu = 600$\,eV) and hard x-ray ($h\nu = 6000$\,eV) VB PES spectra of UCd$_{11}$. In the soft x-ray spectrum, a prominent feature appears near the Fermi energy $E_F$, exhibiting fine structures labeled $A$, $B$, and $C$. The sharp feature $A$ is resolved, thanks to a 60\,meV energy resolution and a fine step width of 10\,meV in the measurement. 
    The data with 1200\,eV incident energy and resolution of 120\,meV (not shown) only exhibit a kink at this energy in the rising part of the spectrum, indicating that a good resolution is needed to resolve feature $A$. 
    Apart from these features, the spectrum is largely featureless. At 600\,eV, the photoionization cross-section of the U\,5$f$ electrons is more than 10 times larger than that of the non-5$f$ states, so that the soft x-ray spectrum predominantly reflects U\,5$f$ contributions. At 6000\,eV, the cross-section of the U\,5$f$ shell is strongly reduced with respect to the other shells. Accordingly, the suppression of features $A$,  $B$ and $C$ in the HAXPES data indicates that these structures originate from the U\,5$f$ states.
	
	In Fig.\,\ref{data}(c), we show the theoretical VB spectra calculated for the two incident photon energies. The orbital-resolved DFT\,+\,DMFT VB intensities of the U\,6$d$, U\,7$s$, U\,7$p$, Cd\,5$s$, and Cd\,5$p$ subshells were obtained using the optimized $\mu_{\rm dc}$ parameter ($\mu_{\rm dc}=5.45$~eV). These spectra were then multiplied by the Fermi function, broadened to account for the experimental resolution, and weighted by the respective tabulated photoionization cross-sections, $\sigma$(600\,eV) and $\sigma$(6000\,eV), given in Refs.~\cite{Trzhaskovskaya2001,Trzhaskovskaya2002,TRZHASKOVSKAYA2018}, to enable a direct comparison with the experiment. Since the cross-sections for U\,7$p$ are not tabulated, those for U\,7$s$ were used instead, based on the empirical observation that the cross-sections per electron for the respective $n$\textit{s} and $n$\textit{p} shells (with $n$ the principal number) are approximately equal. Likewise, because no cross-sections are available for Cd\,5$p$, the values for the next heavier element, In\,5$p$, were adopted. The cross-sections used in this study are listed in Table~I. Here we note that the cross-sections for the Cd\,5$s$ and Cd\,5$p$ need to be increased by a factor of three to obtain a good agreement with the experimental data in the 1.5--8\,eV binding energy range. We infer that the tabulated cross-sections for Cd\,5$s$ and Cd\,5$p$ orbitals may not be accurate for Cd in an alloy, since they were calculated using atomic radial wave functions. The deviations may be substantial for these extended orbitals where the wave functions can change significantly in the alloy environment.

	
	\begin{table}[t]
    \caption{Photoionization cross sections per electron $\sigma_{\rm tot}(h\nu)$ for $h\nu$\,=\,600 and 6000\,eV, calculated following Eq.\,(2) in Ref.\,\cite{Takegami2019}, using the tabulated values in Refs.~\cite{Trzhaskovskaya2001,Trzhaskovskaya2002,TRZHASKOVSKAYA2018}. The tabulated cross sections of Cd\,5$s$ and Cd\,5$p$ are multiplied by a factor 3, see text. All values are given in kb ($10^{-25}$\,m$^2$). }
		\label{tab_S1}
		\setlength{\tabcolsep}{14pt} 
		\begin{tabular}{ c | c c }
			state & $\sigma_{\rm tot}$(600\,eV) & $\sigma_{\rm tot}$(6000\,eV) \\
			\hline \hline
			U\,5$f$ & 28.50  & 0.075 \\
			U\,6$d$ &  1.83 & 0.070 \\
			U\,7$s$ &  0.94 & 0.025 \\
			U\,7$p$ &  0.94 & 0.025 \\		
			Cd\,5$s$&  4.11 x 3 & 0.057 x 3 \\
			Cd\,5$p$&  3.42 x 3 & 0.025 x 3\\
		\end{tabular}
	\end{table}
	
	\begin{figure}[t]
		\begin{center}
			\includegraphics[width=0.85\columnwidth]{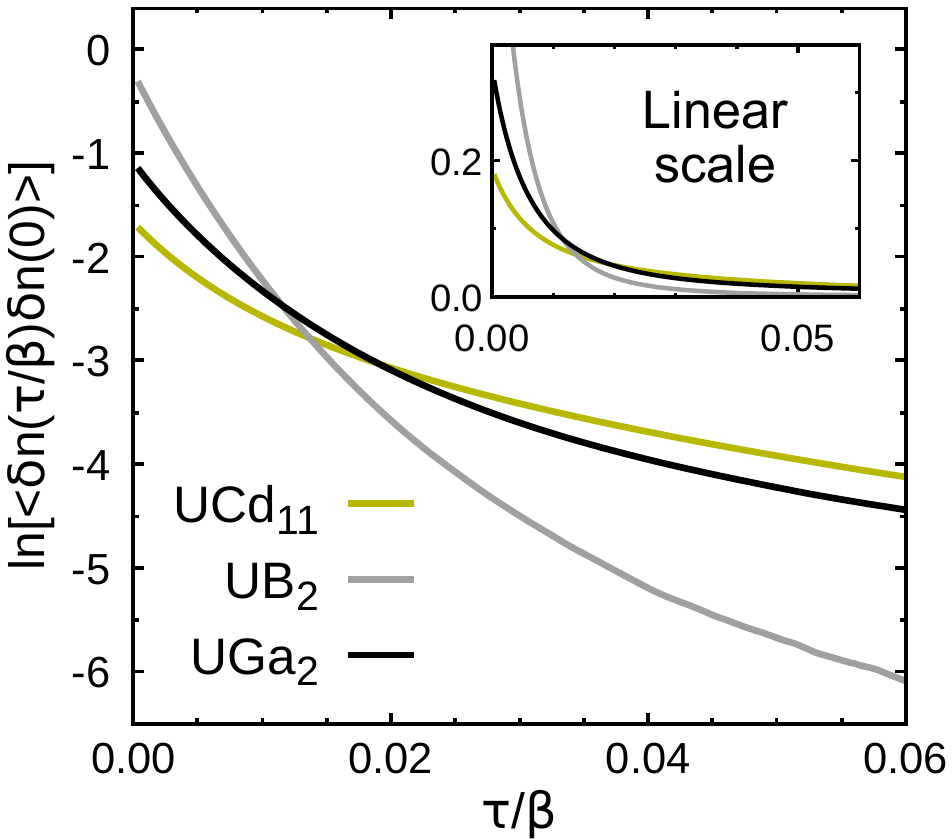}
		\end{center}
    \caption{Logarithm of the charge correlation function $\langle \delta n(\tau) \delta n(0)\rangle$ obtained from DFT\,+\,DMFT calculations for UCd$_{11}$ and the two model compounds UGa$_2$ and UB$_2$.
    Here, $\beta=1/T$ stands for the inverse temperature. Calculations
    were performed at $T= 300$~K. The data for UGa$_2$ and UB$_2$ are adapted from Ref.~\cite{Marino2024}.
    The inset shows a linear-scale plot.}
		\label{charge_chi}
	\end{figure}
    
	\begin{figure}[t]
		\begin{center}
			\includegraphics[width=0.85\columnwidth]{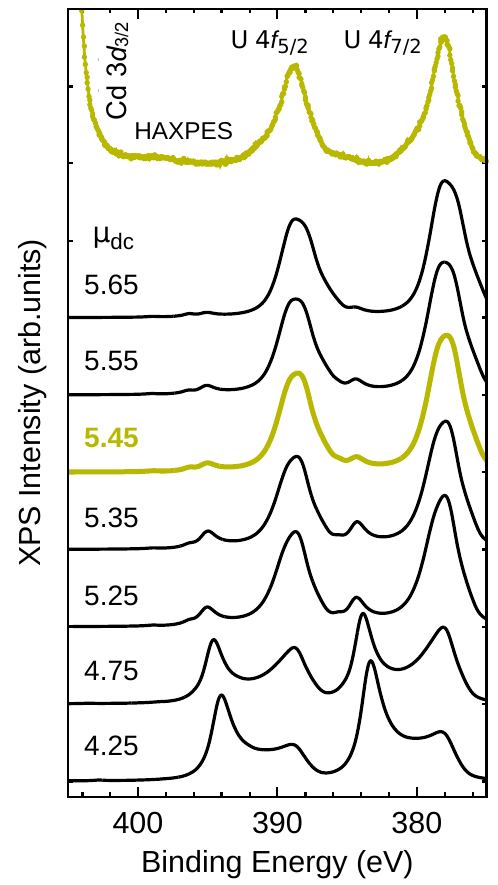}
		\end{center}
		\caption{Experimental U\,4$f$ core-level spectra of UCd$_{11}$ (top), compared to DFT\,+\,DMFT AIM core-level calculations for several values of $\mu_{\rm dc}$. The simulation with the optimum $\mu_{\rm dc}$\,=\,5.45\,eV agrees well with the experimental data. The small feature in the UCd$_{11}$ data at 398\,eV is due to a plasmon excitation (see Appendix).}  
		\label{core_level_mudc}
	\end{figure}
	
	Next, we discuss the electronic model and the uranium valency in UCd$_{11}$. As mentioned in Sec.~II and emphasized in Refs.~\cite{Marino2024,SundermannUTe2} for selected uranium compounds, the double-counting parameter $\mu_{\rm dc}$ is the key factor determining the valency within the DFT\,+\,DMFT approach. Figure~\ref{vb_mu_dep}\,(a) shows the calculated spectral functions for a wide range of $\mu_{\rm dc}$ values, 
    covering extreme reference cases in which the $f^2$ and $f^3$ contributions become nearly saturated at $\mu_{\rm dc}=4.25$\,eV and 6.25\,eV, respectively, although the material does not realize either of them.
	
	The low-energy part of the U\,5$f$ partial density of states in Fig.~\ref{vb_mu_dep}\,(a) is dominated by the $j$\,=\,5/2 spin-orbit states. For small $\mu_{\rm dc}$ ($\mu_{\rm dc}=4.25$--4.75~eV), corresponding to an $f^2$ situation, a metallic quasiparticle peak appears at $E_F$. With increasing $\mu_{\rm dc}$, the $f^3$ contribution grows and the low-energy spectrum develops pronounced multiple substructures. This evolution is accompanied by a substantial redistribution of spectral weight also in the unoccupied region:~a broad feature, indicated by a pink arrow in Fig.~\ref{vb_mu_dep}\,(a), shifts upward in energy as the $f^3$ contribution increases. This feature corresponds to the upper Hubbard band, a hallmark of strong electronic correlation, and becomes most prominent at $\mu_{\rm dc}=6.25$\,eV, where the system approaches the $f^3$ limit and a Mott–Hubbard gap opens at $E_F$.

	Figure~\ref{vb_mu_dep}(b) shows the spectral functions	near the $\mu_{\rm dc}$ value at which the 5$f^3$ configuration becomes more pronounced than 5$f^2$. To resolve this regime, we performed calculations for smaller $\mu_{\rm dc}$ intervals, i.e., from 5.25 to 5.65\,eV. A sharp, spark-like lower branch emerges close to $E_F$, giving rise to feature $A$ observed in the experimental spectrum in Fig.~\ref{data}(a). We find that $\mu_{\rm dc}$ values between 5.25 and 5.45 eV reproduce the experimental features most accurately. Within this optimal range, the upper Hubbard band is clearly present as indicated by an arrow in Fig.~\ref{vb_mu_dep}\,(b), evidencing the strong electronic correlations in UCd$_{11}$.

	In Fig.~\ref{hist}(a)-~\ref{hist}(c), the U 5$f$ valence histogram 
    from
    the DFT+DMFT-optimized model of UCd$_{11}$ (with $\mu_{\rm dc}=5.45$ eV) is shown together with the corresponding histograms of the reference compounds UB$_2$ and UGa$_2$, the latter taken from Ref.~\cite{Marino2024}. In UCd$_{11}$, the 5$f^3$ configuration has by far the highest weight, followed by 5$f^2$ and a minor contribution from 5$f^4$. The weight of other 5$f^n$ configurations is negligible. This narrow distribution across the configurations characterizes UCd$_{11}$ as a strongly correlated U compound, with an average 5$f$ occupation of $\langle n_f\rangle$\,=\,2.87.

   This corresponds to a relatively large filling of the 5$f$ shell, close to the integer configuration 5$f^3$. To further understand this result, we compare the evolution of the 5$f^n$ configurations as a function of $\mu_{\rm dc}$, as it provides insight to the sensitivity of the uranium valence to variations in the U\,5$f$ level energies. This comparison is shown for the three compounds, UCd$_{11}$, UGa$_2$, and UB$_2$, in Figs.~\ref{hist}(d)-~\ref{hist}(f). Only the optimal $\mu_{\rm dc}$ values, indicated by vertical lines in the respective figures, correspond to the physical material. Nevertheless, the comparison reveals that in UCd$_{11}$, the transition between 5$f^2$ and 5$f^3$ occurs within a much narrower $\mu_{\rm dc}$ window than in UGa$_2$, and even more so than in UB$_2$. This indicates that the U\,5$f$ electrons in UCd$_{11}$ are the most strongly localized, i.e., its bands are the narrowest and shifts in the U\,5$f$ level cannot be compensated by hybridization.

    Finally, Fig.~\ref{charge_chi} shows the local charge correlation functions $\langle \delta n(\tau) \delta n(0)\rangle$ obtained from the DFT\,+\,DMFT calculations, which were recently used to examine the degree of localization of the U 5$f$ electrons~\cite{Marino2024,SundermannUTe2}. We show the logarithm of $\langle \delta n(\tau) \delta n(0)\rangle$ in the imaginary-time domain of UCd$_{11}$, together with those of UGa$_2$ and UB$_2$~\cite{Marino2024} for comparison. The amplitude of the local charge fluctuation is encoded in the instantaneous ($\tau=0$) value, which for UCd$_{11}$ is the smallest among the three compounds. This is consistent with the narrow histogram of Fig.~\ref{hist}. Slower decay of $\langle \delta n(\tau) \delta n(0)\rangle$ in UCd$_{11}$compared to UB$_2$, an itinerant system,  as well a UGa$_2$, a intermediate-valence system, reflects a long life time of the charge states due to weak hybridization of the U $5f$ states    with their environment.

	\subsection{Core-level PES with DFT\,+\,DMFT AIM}
	We now turn to the core-level PES data of UCd$_{11}$. Figure~\ref{core_level} compares the U\,4$f$ core-level PES spectrum of UCd$_{11}$ with the calculated spectrum obtained using the DFT\,+\,DMFT AIM. The core-level spectra provide another constraint to the choice of the parameter values, as both the relative weight and the position of the main line and satellite have to be reproduced. Good agreement with the experimental data is achieved only around the optimal $\mu_{\rm dc}$ derived from the VB analysis above (see Fig.~\ref{core_level_mudc}).
	
	We applied the same spectral broadening of 0.5~eV half width at half maximum and aligned the main line to the experimental data. The calculations successfully reproduce the broad  4$f$ core-level emission lines of UCd$_{11}$. 
    This demonstrates that the large width of the U\,4$f$ lines in UCd$_{11}$, compared to those in UB$_2$ (Fig.\,\ref{core_level}), is an intrinsic feature. 
    The difference follows the typical trend observed in core-level spectra from itinerant metallic systems to more localized ones. In itinerant metallic systems, enhanced spectral weight near the threshold due to metallic screening can lead to an asymmetric and comparatively narrow line shape. However, in more localized systems, metallic screening is suppressed and intra-atomic multiplet interactions become more prominent, leading to a broader line shape. This trend is observed, for example, when comparing elemental Fe metal with FeO, Fe$_2$O$_3$, and Fe$_3$O$_4$, where the core-level line shape of the itinerant metal is significantly narrower than those of the more localized oxides~\cite{Liu2014}. The broader main line in UCd$_{11}$ reflects the broad multiplet structure of the $\underline{c}f^3$ ($\underline{c}$ denotes the core hole) final state and is consistent with the strongly localized character of the system. In contrast, in itinerant UB$_2$, the spectrum exhibits an asymmetric narrower main peak.

	
	We would like to note that the observed energy shift of the emission lines in UCd$_{11}$ with respect to UB$_2$ and UGa$_2$ is, as expected, not reproduced, since only relative energies are calculated.
    The slightly broader line shape in the theoretical spectrum may be related to the simplified treatment of the core--valence interaction in the present implementation, or to neglected configuration-dependent lifetime effects. Next, we address the material dependence of the satellites across uranium compounds. In Fig.~\ref{core_level}, the position and low weight of the satellites in UCd$_{11}$ resemble those of UB$_2$. This is counter intuitive, as UB$_2$ is a 5$f^2$-dominant, strongly itinerant system, essentially the opposite limit of UCd$_{11}$, which is a 5$f^3$-dominant, strongly localized system. A prominent satellite, often observed in correlated uranium compounds (e.g., in UGa$_2$, see Fig.~\ref{core_level}), is notably absent in UCd$_{11}$. This demonstrates that the satellite intensity in U\,4$f$ core-level PES spectra cannot be taken as direct evidence of itinerant behavior.

\begin{figure*}[]
		\begin{center}
			\includegraphics[width=2.0\columnwidth]{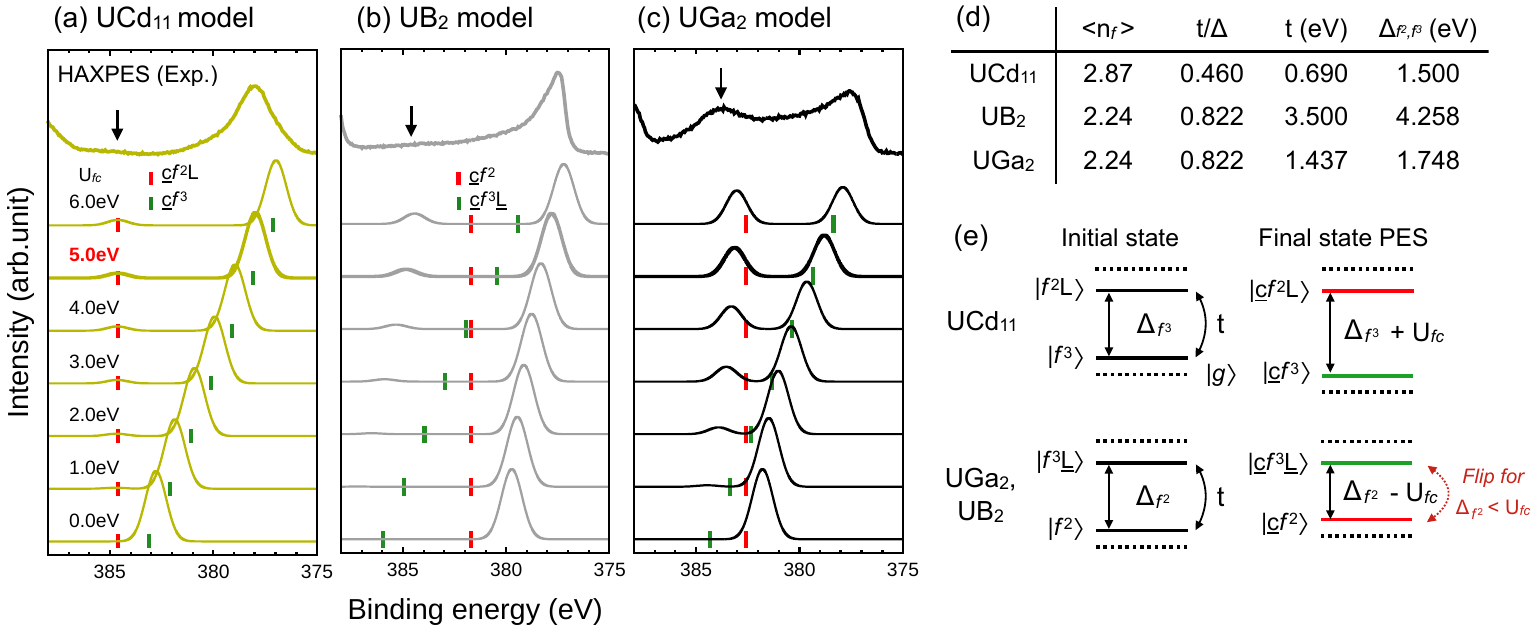}
		\end{center}
    \caption{Two-level model simulation of the core-level PES spectra: (a) UCd$_{11}$ model, (b) UB$_2$ model, and (c) UGa$_2$ model. The spectra are calculated for different values of the core-hole potential $U_{\rm fc}$. The red and green vertical bars indicate the energies of the $|\underline{c}f^2\rangle$ ($|\underline{c}f^3\rangle$) and $|\underline{c}f^3\underline{L}\rangle$ ($|\underline{c}f^3L\rangle$) configurations (with a core hole $c$), respectively, when the hybridization between the two configurations is turned off. (d) Model parameters used in the simulations, and (e) schematic energy diagrams for the initial state and the PES final states. In this model, $\Delta_{f^2}$ ($\Delta_{f^3}$) corresponds to the $f^2$ ($f^3$) formal valence state. In the former case, when $\Delta_{f^2}- U_{\rm fc}$ becomes negative, the energy ordering of the $|\underline{c}f^2\rangle$ and $|\underline{c}f^3\underline{L}\rangle$ configurations is reversed in the PES final state, which occurs in UGa$_2$ and UB$_2$.    }
	\label{toy_model}
\end{figure*}
\subsection{Interpretation of U\,4\textit{f} core-level PES spectra}

The DFT+DMFT AIM approach accurately captures the charge dynamics following the core-level excitation, successfully reproducing the core-level spectra of UCd$_{11}$, UGa$_2$, and UB$_2$, as shown in Fig.~\ref{core_level}. However, due to its complexity the behavior of the numerical model is not readily tractable. Therefore, we adopt a much simplified model based on Ref.~\cite{Marino2024} to illustrate the origin of the weak satellite in UCd$_{11}$ (see Fig.~\ref{toy_model}). 

The two-level model describes the lowest-energy local $f$ configuration and  its dominant charge fluctuation channel. We analyze two cases of either 5$f^2$ or the 5$f^3$ being the lowest-energy state. In case of the low-valence 5$f^2$ being the lowest-energy state, as in UB$_2$ and UGa$_2$, the dominant fluctuation is $|f^2\rangle\leftrightarrow|f^3\underline{L}\rangle$, while for the high-valence 5$f^3$, corresponding to UCd$_{11}$, the dominant fluctuation is $|f^3\rangle\leftrightarrow|f^2L\rangle$ as can be seen in Fig.~\ref{hist}(a) and ~\ref{hist}(b). Here $\underline{L}$ denotes a hole and $L$ an electron in the VB continuum.

When the $|f^2\rangle$ configuration is the lowest-energy state, i.e., in UB$_2$ and UGa$_2$, the model represents the $|f^2\rangle$ and $|f^3\underline{L}\rangle$ states, as illustrated in Fig.~\ref{toy_model}(e). The two states are separated by $\Delta_{f^2}(>0)$ and coupled via hybridization $t$, and are described by the Hamiltonian (in the basis of $|f^2\rangle$ and $|f^3\underline{L}\rangle$),
\begin{equation*}
	\hat{H}_{f^2} =
	\begin{pmatrix}
		0 & t \\
		t & \Delta_{f^2}
	\end{pmatrix}.
\end{equation*}
In the PES final states, the core-hole potential $U_{\rm fc}(>0)$ modifies the Hamiltonian for $|\underline{c}f^2\rangle$ and $|\underline{c}f^3\underline{L}\rangle$ as
\begin{equation*}
	\hat{H}_{{\rm PES},f^2} =
	\begin{pmatrix}
		0 & t \\
		t & \Delta_{f^2} - U_{\rm fc}
	\end{pmatrix}.
\end{equation*}
The potential $U_{\rm fc}$ affects the $|\underline{c}f^2\rangle$ and $|\underline{c}f^3\underline{L}\rangle$ states differently because they contain different numbers of $f$ electrons, thereby altering the energy separation in the final state, as shown in Fig.~\ref{toy_model}(e).

In UCd$_{11}$, the $|f^3\rangle$ configuration constitutes the lowest-energy state. Accordingly, the two levels represent $|f^3\rangle$ and $|f^2L\rangle$, where $|L\rangle$ denotes an electron in the VB continuum. With the energy separation $\Delta_{f^3} (> 0)$ with the $|f^2L\rangle$ state [see Fig.~\ref{toy_model}\,(e)], $\hat{H}_{f^3}$ reads
\begin{equation*}
	\hat{H}_{f^3} =
	\begin{pmatrix}
		0 & t \\
		t & \Delta_{f^3}
	\end{pmatrix},
\end{equation*}
for the basis $|f^3\rangle$ and $|f^2L\rangle$. In the PES final states ($|\underline{c}f^3\rangle$ and $|\underline{c}f^2L\rangle$),
\begin{equation*}
	\hat{H}_{{\rm PES},f^3} =
	\begin{pmatrix}
		0 & t \\
		t & \Delta_{f^3} + U_{\rm fc}
	\end{pmatrix},
\end{equation*}
where $|\underline{c}f^2L\rangle$ is shifted by $U_{\rm fc}$. 

In the present model the ground states, $|g\rangle = \alpha |f^2\rangle + \beta |f^3\underline{L}\rangle$ or $|g\rangle = \alpha |f^2L\rangle + \beta |f^3\rangle$, is determined by the ratio $t/\Delta_{f^2,f^3}$. For UB$_2$ and UGa$_2$, we use the parameters from Ref.~\cite{Marino2024}, which are summarized in Fig.~\ref{toy_model}\,(d): $t/\Delta_{f^2} = 0.822$ with $t = 3.5$\,eV for UB$_2$ and $t = 1.437$\,eV for UGa$_2$, yielding an average 5$f$ occupation of $n_{5f} = 2.24$, i.e., dominated by the $f^2$ configuration. For UCd$_{11}$, we use $t/\Delta_{f^3}$\,=\,0.460 with $t$\,=\,0.690\,eV, yielding $n_{5f}$\,=\,2.87, consistent with the DFT\,+\,DMFT result in Sec.~B. 

Figure~\ref{toy_model}(a)-~\ref{toy_model}(c) show the PES spectra of these models for different vaules of $U_{\rm fc}$ values. For the $f^2$ compounds, UGa$_2$ and UB$_2$, the energy difference between the two configurations in the PES final state evolves as $\Delta_{f^2} - U_{\rm fc}$, see the red and green ticks in Figs.~\ref{toy_model}(b) and ~\ref{toy_model}(c) for $|\underline{c}f^2\rangle$ and $|\underline{c}f^3\underline{L}\rangle$. As a result, the energies of the two configurations are inverted when $U_{\rm fc}$ exceeds $\Delta_{f^2}$. This inversion does not occur in UCd$_{11}$ owing to its formal $f^3$ valence, where the energy separation from $|f^2L\rangle$ is given by $\Delta_{f^3} + U_{\rm fc}$. Thus, the energy separation between $|\underline{c}f^2L\rangle$ (red ticks) and $|\underline{c}f^3\rangle$ (green ticks) increases monotonically, as shown in Fig.~\ref{toy_model}(a).

The simple models, implementing $U_{\rm fc}=5.0$ eV, the same value as in the DFT+DMFT AIM, capture the experimental data nicely. Despite clear differences in the initial state occupation $n_f$ and hybridization $t$ between UCd$_{11}$ and UB$_2$, the resulting spectra closely resemble each other. The weak satellite in the UB$_2$ model originates from the larger hybridization $t$; reducing $t$ produces a more pronounced satellite~\cite{Marino2024}, as for UGa$_2$. The weaker satellite in UCd$_{11}$, on the other hand, arises predominantly from the absence of the energy-level crossing between the $f^3$ and $f^2$ configurations in the PES final states. 

\subsection{Implications for UCd$_{11}$}
UCd$_{11}$ is identified as a strongly localized uranium compound with a predominant $5f^{3}$ configuration and an average occupation of $\langle n_{f}\rangle$\,$\approx$\,2.87. These results are consistent with a localized Kramers doublet ground state, in agreement with specific-heat\,\cite{Yamamoto2012} and de Haas–van Alphen\,\cite{Hirose2013} studies. They further imply that the magnetism in UCd$_{11}$ arises from local moments rather than from induced moments due to coupling of a singlet ground state to excited crystal-field states as in UGa$_2$\,\cite{Marino2023} or Fe-doped URu$_2$Si$_2$\,\cite{Marino2023a}. 

Furthermore, low-temperature neutron diffraction measurements are consistent with a large U\,5$f$ ordered moment, though were unable to find evidence for magnetic peaks associated with a finite-Q wavevector and antiferromagnetism\,\cite{Thompson1988}.  Our result of a localized 5$f^3$ configuration provide further motivation to investigate the nature of the magnetically ordered state in UCd$_{11}$.
    
\section{Summary}
We have developed an electronic structure model for UCd$_{11}$ based on the DFT\,+\,DMFT method, using material-specific parameters tuned to reproduce experimental valence-band photoemission spectroscopy data. Our calculations show that the U\,5\textit{f} electrons in UCd$_{11}$ are strongly localized and adopt the U $5f^3$ configuration in agreement with inelastic x-ray scattering and x-ray absorption results. The U\,4$f$ core-level spectra of UCd$_{11}$ are successfully reproduced using the DFT\,+\,DMFT Anderson impurity model, including the weak satellite feature that has long been a source of controversy regarding the degree of U\,$5f$ localization inferred from different spectroscopic techniques.
	
By combining DFT\,+\,DMFT simulations with a simple model analysis across UCd$_{11}$ ($f^{3}$-dominant, localized), UB$_2$ ($f^{2}$-dominant, itinerant), and UGa$_2$ ($f^{2}$-dominant, localized), we provide a comprehensive picture of the U\,4$f$ core-level spectra as a function of dominant U\,5\textit{f} configuration and varying degrees of localization. In UCd$_{11}$, the weak satellite intensity does not signal itinerant U 5$f$ behavior, but rather reflects a stable localized 5$f^{3}$ configuration. This demonstrates that the satellite strength in the core-level spectrum is not a universal indicator of itinerancy.

\section{Acknowledgment}
All authors acknowledge DESY (Hamburg, Germany), a member of the Helmholtz Association HGF, for the provision of experimental facilities, and thank Gertrud Zwicknagl and Peter Thalmeier for enlightening discussions. A.H. acknowledges K.-H.Ahn for helpful discussions and ~was supported by JSPS KAKENHI Grant No.25K00961, No.25K07211, No.23H03816, No.23H03817, and No.23K03324, and the 2025 Osaka Metropolitan University (OMU) Strategic Research Promotion Project (Young Researcher). S.-i.F. was funded by JSPS KAKENHI Grant No.18K03553, No.20KK0061 and No.22H03874. Parts of the computations were performed at It4innovations funded by the Ministry of Education, Youth and Sports of the Czech Republic through the e-INFRA CZ (ID:90254). A.S. acknowledges support from the German Research Foundation (DFG) - grant No.387555779 and No.567326535.  Work at Los Alamos National Laboratory was performed under the auspices of the U.S. Department of Energy, Office of Basic Energy Sciences, Division of Materials Science and Engineering under project “Quantum Fluctuations in Narrow-Band Systems”. Part of the computations in this work were performed using the facilities of the Supercomputer Center, the Institute for Solid State Physics, the University of Tokyo.

\begin{figure*}[t]
	\begin{center}
		\includegraphics[width=1.7\columnwidth]{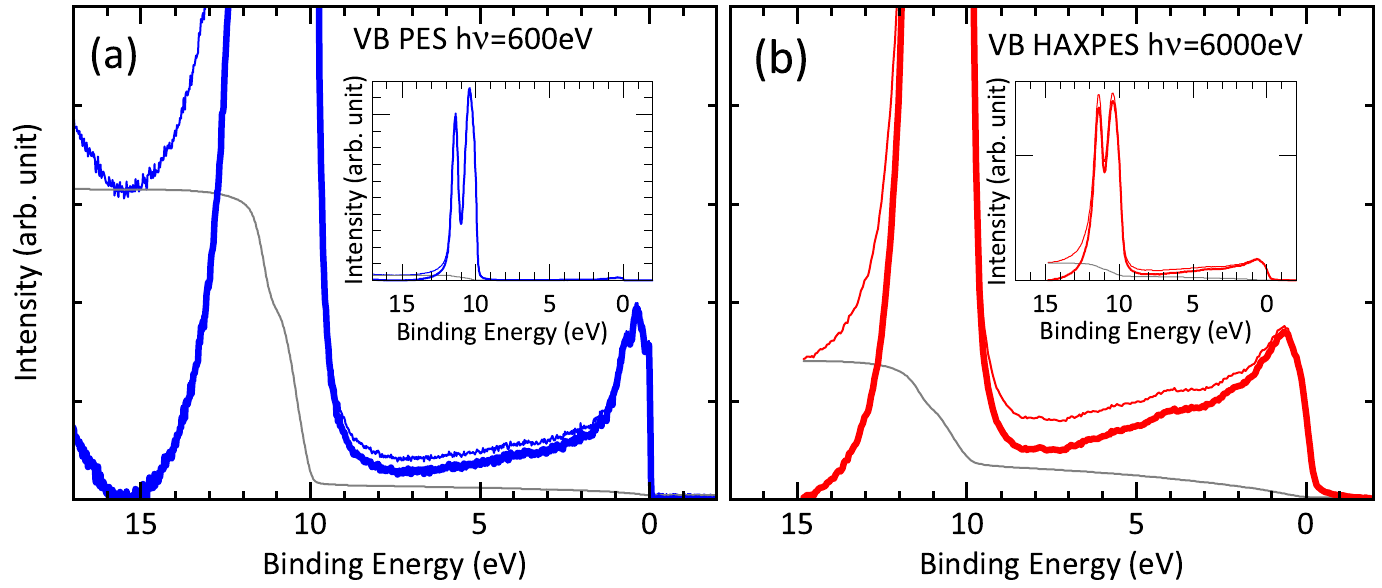}
	\end{center}
	\caption{Valence band and Cd\,4$d$ core-level PES spectra of UCd$_{11}$ for 600\,eV [panel(a)] and 6000\,eV [panel(b)] incident energy. The respective thin colored lines represent the spectra as measured and the thick colored lines after subtraction of an integral-type (Shirley) background (grey lines). }
	\label{BG_corr}
\end{figure*}

\begin{figure}[h]
	\begin{center}
		\includegraphics[width=0.99\columnwidth]{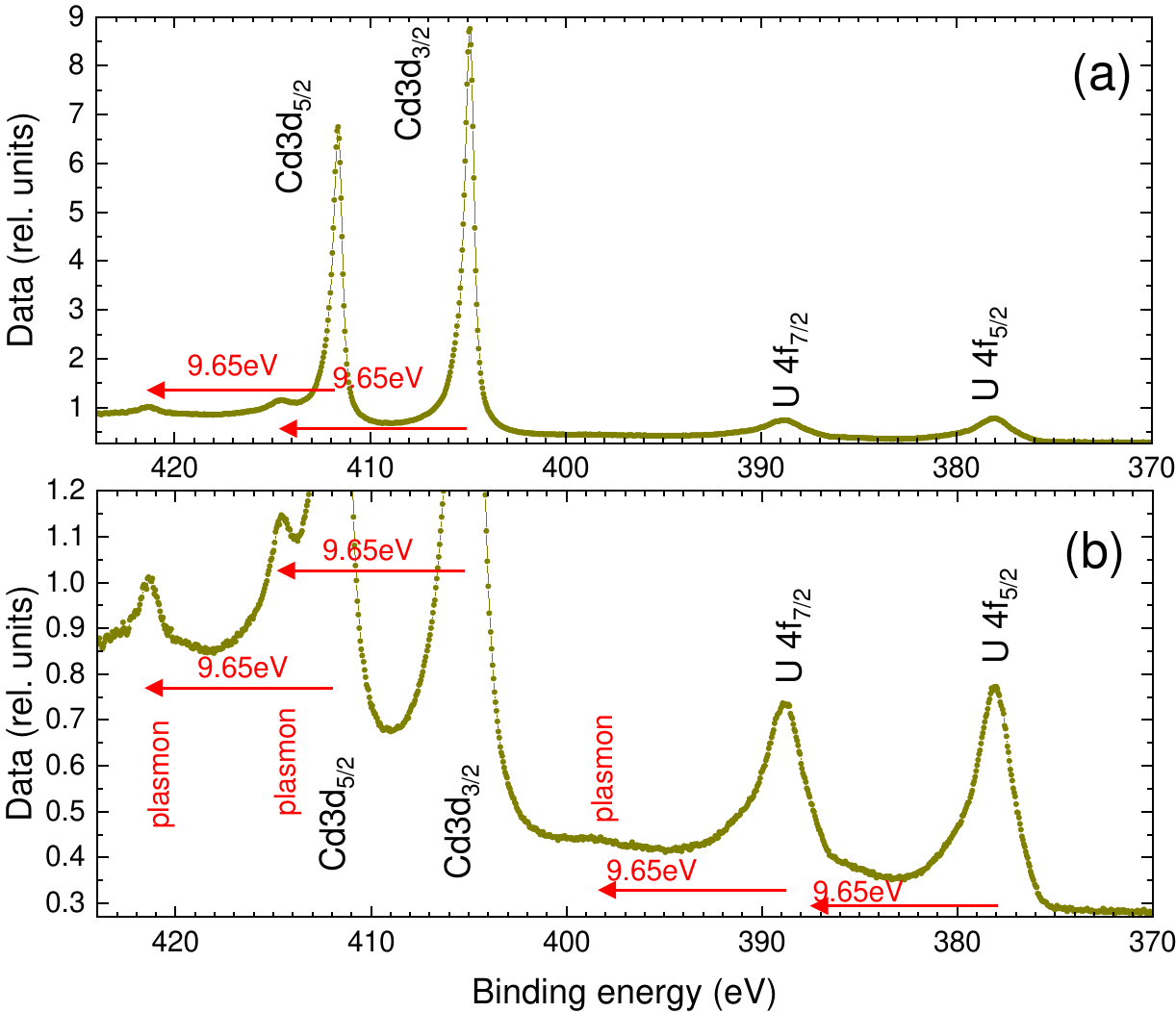}
		\includegraphics[width=0.99\columnwidth]{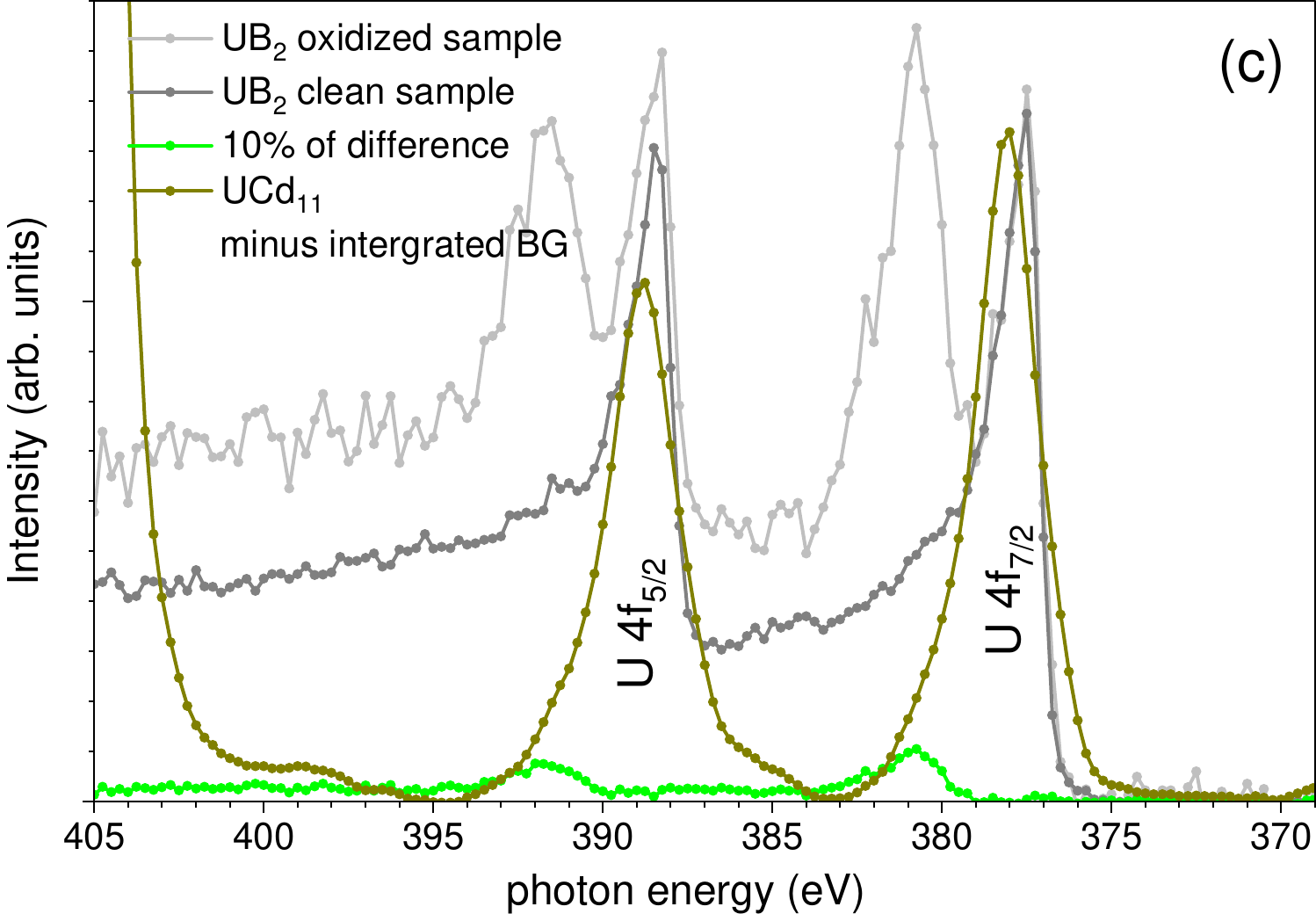}
	\end{center}
	\caption{Panels (a) and (b) show the same data but with different intensity scales. The red arrows point out the energy position of the plasmons. Panel (c) shows the U 4$f$ emission lines of oxidized UB$_2$ (light gray), clean UB$_2$ (dark gray) and 10\% of the difference plot (green line) as well as integral-type background (Shirley) corrected UCd$_{11}$ (dark yellow). }
	\label{plasmon}
\end{figure}

\section{Supplemental data}
The data supporting this study's findings are available within the article and can be made available in digital form upon reasonable request.
\appendix

\section{Supplemental data}
\label{appendix:exp}

Figure\,\ref{BG_corr} shows the valence band and the shallow Cd\,4$d$ core level PES spectra, taken with 600\,eV [Fig.~\ref{BG_corr}(a)] and 6000 eV [Fig.~\ref{BG_corr}(b)] incident photon energy. An integral-type ("Shirley") background correction has been applied such that the intensity at the higher binding energy side of the Cd\,4$d$ is set to zero. Thin lines are the spectra as measured and thick lines after the background correction. There is appreciable intensity in the valence band region between 1.5\,eV and 8\,eV binding energy, and this is mainly due to the presence of the Cd\,5$s$ and 5$p$ states as reproduced by the DFT\,+\,DMFT calculations [see Fig.\,\ref{data}\,(c),\,\ref{data}(d) and Fig.\,\ref{vb_mu_dep}\,(a)].

Figure\,\ref{plasmon}\,(a) and \ref{plasmon}\,(b) shows the Cd\,3$d$ and U\,4$f$ emission lines, respectively, with Fig.\,\ref{plasmon}\,(b) plotted using a stretched $y$-scale to enhance the U\,4$f$ intensities. These two panels demonstrate that the feature at 398\,eV in the U\,4$f$ core-level spectrum of UCd$_{11}$ is due to a plasmon excitation. It appears at the same energy offset, namely 9.65\,eV away form the main emission line, as the plasmon peaks observed for the Cd\,3$d$ emission lines. 
	

Figure\,\ref{plasmon}\,(c) highlights the U\,4$f$ core-level emission lines. Here, the spectra of  oxidized UB$_2$ (light grey) and clean UB$_2$ (dark grey) are compared. The spectrum of the oxidized sample exhibit the characteristic additional features approximately 3.5\,eV above the main emission lines of the clean sample. Hence, the difference of the two spectra, scaled to 10\%  (green line), reflects the contribution from a uranium oxide layer. This comparison demonstrates that any oxide-related contributions in the UCd$_{11}$ data must be very small, certainly too small for oxide-layer satellites features to be visible in the spectra.

In Eq.~\eqref{eq:localH}, the local Hamiltonian at the U site, constructed via numerical Wannierization of the DFT bands, is expressed in the crystal-field basis. The orbital ordering is
$\{ z(5z^2-3r^2),\, y(5y^2-r^2),\, x(5x^2-r^2),\, z(x^2-y^2),\, xyz,\, y(y^2-3x^2),\, x(x^2-3y^2) \}$
in each spin sector (up and down).

\begin{widetext}
\begin{equation}
H_{\rm loc} =
{\scriptsize
\begin{pmatrix}
0.667 & 0 & 0 & 0 & 0 & 0 & 0 & 0 & -0.314 & -0.314\,\mathrm{i} & 0 & 0 & 0 & 0 \\
0 & 0.636 & 0.129\,\mathrm{i} & 0 & 0 & 0.007 & 0 & 0.314 & 0 & 0 & 0.203 & -0.202\,\mathrm{i} & 0 & 0 \\
0 & -0.129\,\mathrm{i} & 0.636 & 0 & 0 & 0 & 0.007 & 0.314\,\mathrm{i} & 0 & 0 & -0.203\,\mathrm{i} & -0.202 & 0 & 0 \\
0 & 0 & 0 & 0.692 & -0.255\,\mathrm{i} & 0 & 0 & 0 & -0.203 & 0.203\,\mathrm{i} & 0 & 0 & 0.158 & -0.158\,\mathrm{i} \\
0 & 0 & 0 & 0.255\,\mathrm{i} & 0.635 & 0 & 0 & 0 & 0.202\,\mathrm{i} & 0.202 & 0 & 0 & 0.156\,\mathrm{i} & 0.156 \\
0 & 0.007 & 0 & 0 & 0 & 0.709 & -0.383\,\mathrm{i} & 0 & 0 & 0 & -0.158 & -0.156\,\mathrm{i} & 0 & 0 \\
0 & 0 & 0.007 & 0 & 0 & 0.383\,\mathrm{i} & 0.709 & 0 & 0 & 0 & 0.158\,\mathrm{i} & -0.156 & 0 & 0 \\
0 & 0.314 & -0.314\,\mathrm{i} & 0 & 0 & 0 & 0 & 0.667 & 0 & 0 & 0 & 0 & 0 & 0 \\
-0.314 & 0 & 0 & -0.203 & -0.202\,\mathrm{i} & 0 & 0 & 0 & 0.636 & -0.129\,\mathrm{i} & 0 & 0 & 0.007 & 0 \\
0.314\,\mathrm{i} & 0 & 0 & -0.203\,\mathrm{i} & 0.202 & 0 & 0 & 0 & 0.129\,\mathrm{i} & 0.636 & 0 & 0 & 0 & 0.007 \\
0 & 0.203 & 0.203\,\mathrm{i} & 0 & 0 & -0.158 & -0.158\,\mathrm{i} & 0 & 0 & 0 & 0.692 & 0.255\,\mathrm{i} & 0 & 0 \\
0 & 0.202\,\mathrm{i} & -0.202 & 0 & 0 & 0.156\,\mathrm{i} & -0.156 & 0 & 0 & 0 & -0.255\,\mathrm{i} & 0.635 & 0 & 0 \\
0 & 0 & 0 & 0.158 & -0.156\,\mathrm{i} & 0 & 0 & 0 & 0.007 & 0 & 0 & 0 & 0.709 & 0.383\,\mathrm{i} \\
0 & 0 & 0 & 0.158\,\mathrm{i} & 0.156 & 0 & 0 & 0 & 0 & 0.007 & 0 & 0 & -0.383\,\mathrm{i} & 0.709 \\
\end{pmatrix}
}
\label{eq:localH}
\end{equation}
\end{widetext}

\bibliography{references_UCd11_resub.bib}
\end{document}